\documentclass{ametsocV6.1}

\title{The Role of Sea-ice Processes on the Probability of AMOC Transitions}

\authors{Ren\'e M. van Westen,\aff{a}\correspondingauthor{Ren\'e M. van Westen, r.m.vanwesten@uu.nl} 
Val\'erian Jacques-Dumas, \aff{a}
Amber A. Boot, \aff{a}
Henk A. Dijkstra,\aff{a} 
}

\affiliation{\aff{a}{Institute for Marine and Atmospheric research Utrecht, Department of Physics, 
Utrecht University, Princetonplein 5, Utrecht, 3584 CC, the Netherlands}
}

\abstract{Recent simulations performed with the  Community Earth System Model (CESM) 
have suggested a crucial role of sea-ice processes in AMOC hysteresis behaviour under 
varying surface freshwater forcing. Here, we further investigate this issue using additional
CESM simulations and a novel conceptual ocean-sea-ice box model. The CESM simulations 
show that the presence of sea ice gives rise to the existence of statistical 
equilibrium states with a weak AMOC strength.
This is confirmed in the conceptual model, which captures the same 
AMOC hysteresis behaviour as in the CESM simulation  and  where steady states are computed 
versus forcing parameters. In the conceptual model, transition probabilities between the different 
equilibrium states are determined using rare event techniques.  The transition probabilities 
from a strong AMOC state to a weak AMOC state increase when considering sea-ice 
processes and indicate that sea ice promotes these transitions.  On the other hand,  sea 
ice strongly reduces the probabilities of the reverse transition from a  
weak AMOC state to a strong AMOC state and this  implies that sea ice also limits 
AMOC recovery.  The results here indicate that sea-ice processes play a dominant role 
in AMOC hysteresis width  and influence transition probabilities between the different 
equilibrium states.
} 

\begin{document}

\maketitle

%
%
%
\statement
We develop a novel conceptual ocean-sea-ice  box model to explain AMOC hysteresis behaviour 
recently found  in the Community Earth System Model and determine how sea-ice processes 
influence the hysteresis width and the probability of AMOC transitions.  
%
%

%
\section{Introduction}

The Atlantic Meridional Overturning Circulation (AMOC) plays an important role in regulating the
global climate by redistributing heat and salinity through the global ocean. 
The effects of anthropogenic climate change on the AMOC provide reasons for concern  as the 
AMOC is a tipping element in the climate system \citep{Armstrong2022}.
There is an ongoing debate whether the present-day AMOC is close to its tipping point 
\citep{Boers2021, Ditlevsen2023, Mehling2023}, but it is known that AMOC tipping has severe global 
climate impacts \citep{Orihuela2022, vanWesten2024}.  So far,  historical AMOC reconstructions 
indicate  that its strength is declining since 1900 \citep{Caesar2018, Caesar2021} and  projections 
of climate models indicate that this decline will continue under future climate change and its decline 
is insensitive to the used emission scenario \citep{Weijer2020}. 

The potential tipping of the AMOC has been studied in a hierarchy of models \citep{Dijkstra2023b}, 
from simple box  models  \citep{Stommel1961, Cessi1994, Cimatoribus2014},  ocean-only models   
\citep{Dijkstra2007},  to various Earth system Models of Intermediate  Complexity (EMICs) 
\citep{Rahmstorf2005},  early Global Climate Models (GCMs) \citep{Hawkins2011, Hu2012}, 
and in modern  Earth System Models (ESMs) \citep{Jackson2022}. In box models and 
ocean-only  models, steady state calculations  have shown that multi-stable AMOC regimes 
can exist.  Transitions between strong and collapsed AMOC states are caused by the salt-advection 
feedback, where a freshwater anomaly in the North Atlantic weakens the AMOC strength 
and reduces the associated salinity transport. This response  then amplifies the original 
freshwater  anomaly \citep{Marotzke2000}.   In EMICs and ESMs, an impression of such a 
multi-stable regime can be obtained by slowly  increasing the surface freshwater forcing 
to induce an AMOC collapse.  When  reversing the surface  freshwater forcing,  the 
AMOC often  recovers at smaller values of  this forcing compared to the collapse, giving rise 
to hysteresis behaviour \citep{Rahmstorf2005, Hawkins2011, Hu2012}. 

The presence of a multi-stable AMOC in a hierarchy of models, and possibly in the actual 
climate system, has far reaching consequences. When the present-day AMOC is in such a
regime, there is a finite probability of a noise-induced transition between the different 
equilibrium states, given the presence of stochastic surface forcing. This probability will
increase when approaching the tipping point; in addition, rate-induced transitions may 
occur \cite[]{Ashwin2012, Lohmann2021}.  Transition probabilities between the different 
AMOC states, which may arise under stochastic surface freshwater flux noise,  have so 
far been only studied in idealised models  \citep{Castellana2019, Baars2021}.  
For state-of-the-art climate models determining transition probabilities is problematic 
since this is computationally very expensive, in particular when the transition probabilities are very 
low  \citep{Jacques-Dumas2023}. 

This  study is motivated by recent results where such hysteresis behaviour was found in the 
Coupled Model Intercomparison Project phase 5 (CMIP5) version of the Community Earth 
System Model (CESM)  \citep{vanWesten2023c, 
vanWesten2024}. The transition (`collapse') between the strong (`AMOC on') to the weak 
(`AMOC off')  state was very similar to the one expected from ocean-only models  
\citep{Dijkstra2007}, whereas the reverse transition (`recovery') behaved quite differently. 
It was shown that  sea-ice ocean interactions  play an important role in the AMOC recovery 
because the relatively weak ($<$5~Sv, 1~Sv = 10$^6$~m$^3$~s$^{-1}$)  northward 
overturning cell in the Atlantic Ocean  is strongly suppressed  by the extensive North 
Atlantic sea-ice cover \citep{vanWesten2023c}.   However, the precise effects of the 
sea ice on the multi-stable AMOC regime could not be investigated in CESM, where 
the surface freshwater forcing was transient. 

In this paper, we show that the sea ice actually increases the hysteresis width in the CESM 
and explain the physics of this result using a novel conceptual ocean-sea-ice  model. Using 
this conceptual model, we also  extend the study of  \citet{Castellana2019} by looking at 
transition probabilities  between  the different equilibrium AMOC  states in the  presence 
of sea ice.  The conceptual ocean-sea-ice model is described in Section~2. Next in Section~3, 
we first present additional CESM results and then  analyse the steady states and 
transitions between the steady  states in the conceptual model.
Finally in Section~4, we summarise and discuss the results.  

\section{Models and Methods} 

\subsection{CESM configuration} 

In \citet{vanWesten2023c}, version 1 of the CESM was 
used, with   horizontal resolutions of  1$^{\circ}$  for the  ocean/sea ice  and 2$^{\circ}$ for the 
atmosphere/land components.  The CESM simulation was initialised from the end of a 2,800-year long 
pre-industrial control simulation \citep{Baatsen2020} under constant greenhouse gas, solar 
and aerosol forcings set to pre-industrial levels.     The hysteresis behaviour in the CESM 
\citep{vanWesten2023c} was  obtained by varying the surface freshwater forcing ($F_H$) between 
latitudes 20$^{\circ}$N and 50$^{\circ}$N in the Atlantic Ocean. This freshwater flux anomaly was 
compensated over the rest of the domain to conserve salinity, this a typical set-up for a hosing experiment. 
The surface freshwater forcing linearly increased  in time with a rate of  $3 \times 10^{-4}$~Sv~yr$^{-1}$ up to a maximum of 
$F_H = 0.66$~Sv  in model year~2,200. This rate of change in $F_H$ is  quite small with the 
aim to stay close to  equilibrium, often referred to as a so-called quasi-equilibrium approach 
\citep{Rahmstorf2005, Hu2012}. At the end of the simulation the forcing was reversed and 
$F_H$ decreased at the same rate and reaches zero in model year~4,400.

The different AMOC states in  \citet{vanWesten2023c} are not in statistical equilibrium 
\citep{Hawkins2011},  given the quasi-equilibrium approach, and therefore we performed 
additional CESM simulations under fixed  freshwater forcing values.  From the hysteresis 
simulation we branch off 4~simulations from model years~600 and 3,800 ($F_H = 0.18$~Sv) 
and from model years~1,500 and 2,900 ($F_H = 0.45$~Sv), and keep the value of $F_H$ constant during each simulation. 
The AMOC collapse (recovery) occurs around model year~1,750 (4,100), 
this motivates the branching points from $F_H = 0.45$~Sv) ($F_H = 0.18$~Sv).
The branched simulations are integrated for 500~years and we 
determine their statistical properties over the last 50~years. 

\subsection{A conceptual ocean-sea-ice model}

\begin{figure}
\center
\includegraphics[width=1\columnwidth, trim = {3cm 0cm 0cm 0cm}, clip]{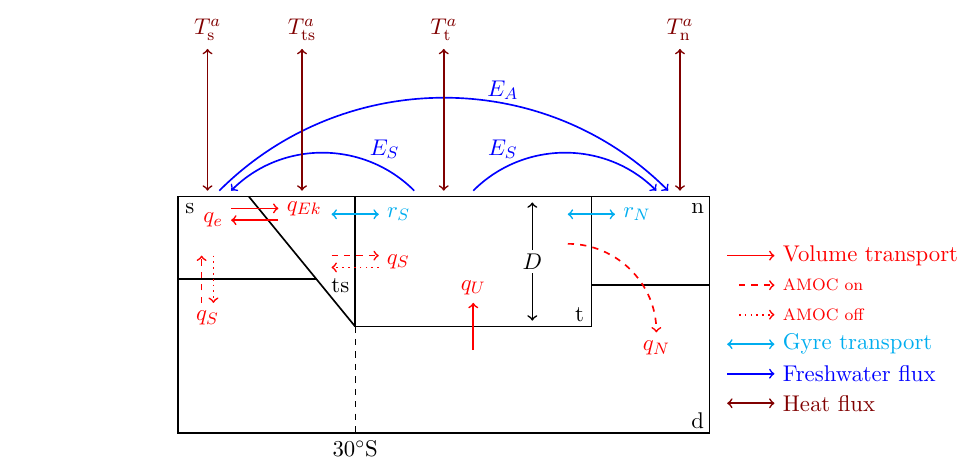}
\caption{Schematic representation of the 5-box AMOC model, adapted from \cite{Cimatoribus2014} and 
\citet{Castellana2019} (the CCM). The red arrows represent the volume fluxes, whereas the dashed 
and dotted arrows indicate the AMOC on and AMOC off states, respectively. The cyan and blue arrows 
are the gyre transport and freshwater fluxes. The brown arrows are the heat fluxes with the overhead 
atmosphere for each surface box (i.e., box~s, ts, t and n).} 
\label{fig:Figure_1}
\end{figure}

The new model is based on  the  ocean box model which was developed by \citet{Cimatoribus2014} 
and extended by  \citet{Castellana2019} (hereafter called CCM, see Figure~\ref{fig:Figure_1}).
The CCM consists of conservation equations for the salinity content of boxes~s, ts, t 
and n, one equation to conserve salinity (from which the salinity of box~d is derived)
and one equation for the time-varying pycnocline depth ($D$). The AMOC strength in the 
northern box ($q_N$) is given by 
\begin{equation} \label{eq:q_n}
q_N       = \eta \frac{\rho_n - \rho_{ts}}{\rho_0} D^2 , 
\end{equation}
where $\eta$ is a downwelling (pumping) constant, $\rho_n - \rho_{ts}$ is the density 
difference between box~n and box~ts and  $\rho_0$ is a reference density. The densities 
are determined from a linear equation of state. 
In the CCM, the temperatures for box~n and box~ts are input parameters (5$^{\circ}$C 
and 10$^{\circ}$C, respectively) and are fixed in time.
The main forcing parameter is the asymmetric freshwater flux ($E_A$) from box~s to box~n.
Note that the hosing set-up in the CCM is slightly different than in the CESM where fresh water is added between 20$^{\circ}$N and 50$^{\circ}$N and compensated elsewhere.
Under large $E_A$,  the density difference $\rho_n - \rho_{ts}$ becomes negative and the circulation 
reverses (red dashed arrows  become red dotted arrows in Figure~\ref{fig:Figure_1}). 
The CCM is extensively discussed in  \citet{Castellana2019} and, as we use it 
for comparison to the new model where sea ice is included,  the governing 
equations and input parameter settings are  provided in the Appendix 
(see equations~\ref{eq:Box_model_salt}). 

Our aim is to couple  sea-ice processes to the CCM and hence we 
first extend it  by including oceanic temperatures.   The governing equations for temperature 
(see equations~\ref{eq:Box_model_temp} in the Appendix) are very similar to the ones for salinity.
The main difference is that the equations for
the surface boxes (box~s, ts, t and n) have one additional term which represents the 
heat exchange with the atmosphere (brown arrows in Figure~\ref{fig:Figure_1}). The change 
in ocean heat content for each surface box~($i$ = s, ts, t and n) and box~d is then given 
by (under the assumption of constant heat capacity and reference density):
\begin{subequations} \label{eq:TEMP_only}
\begin{align}
 \frac{\mathrm{d}(V_i T_i)}{\mathrm{d}t} &=  F_i - \lambda^a A_i (T_i - T_i^a) ,  \\
  \frac{\mathrm{d}(V_\mathrm{d} T_\mathrm{d})}{\mathrm{d}t} &= F_\mathrm{d} . 
\end{align}
\end{subequations}
Here, $F_i$ is the advective oceanic heat transport,  $V_i$ the volume, $A_i$ 
the horizontal surface area,   $\lambda^a$ the heat exchange rate with the 
atmosphere,  $T_i$ the ocean temperature and $T_i^a$ the overhead  
atmospheric temperature. Note that the temperature for box~d is a dynamical
quantity  whereas the salinity of this box is diagnostically derived from  salinity 
conservation. 

The heat exchange rate $\lambda^a$ and atmospheric temperatures are additional parameters  in the  
extended CCM (hereafter E-CCM) that need to be tuned (Appendix and Figure~\ref{fig:Figure_A1}).
For an arbitrary value of $E_A$ (and for the AMOC on state), we tune these parameters such that the 
ocean temperatures for box~n and box~ts are 5$^{\circ}$C and 10$^{\circ}$C, respectively. 
We mainly use $\lambda^a = 3.5 \times 10^{-6}$~m~s$^{-1}$ below and for the given $E_A$
the AMOC on state of the E-CCM is identical to that of the CCM.
We used $E_A = 0.0644$~Sv for tuning and varying this tuning value slightly shifts the bifurcation points in the E-CCM.
In the CCM this is the minimum value of $E_A$ for which the
freshwater  transport carried by  the AMOC at 30$^{\circ}$S, below indicated by 
$F_{\mathrm{ovS}}$, becomes negative  \citep{Castellana2019}. 
The quantity $F_{\mathrm{ovS}}$ is determined from the model as 
\begin{equation} \label{eq:q_s}
F_{\mathrm{ovS}}       = -\frac{q_S}{S_0} \left(S_\mathrm{ts} - S_\mathrm{d} \right) , 
\end{equation}
where $S_0$ is a reference salinity. 
The quantity $F_{\mathrm{ovS}}$ is important as it is a measure for the salt-advection feedback strength.
The role of $F_{\mathrm{ovS}}$ is more versatile in the CCM because when it is 
negative (positive) in the AMOC on state the model is in its 
multi-stable (monostable) regime. 

Finally, the sea-ice processes are represented as follows and specifically for $\lambda^a = 3.5 \times 10^{-6}$~m~s$^{-1}$.  
When the temperature for box~n is lower than 5$^{\circ}$C (its reference temperature), we assume that the sea-ice 
fraction of box~n ($f_\mathrm{n}$) grows linearly by 20\% per degree cooling  (Figure~\ref{fig:Figure_A2}a).  
This is an arbitrary choice for  the sea-ice growth and we do not use the freezing temperature for sea-ice growth as box~n
represents the volume-averaged temperature over the higher latitudes.
To represent now the effect that sea ice suppress upper ocean mixing \citep{Lin2023}, 
we multiply the AMOC strength by a given AMOC reduction factor $0 \leq r_q \leq 1$ such that the 
AMOC strength becomes 
\begin{equation} \label{eq:q_n_ice}
q_N^{\mathrm{ice}} = r_q ~ q_N , 
\end{equation}
where $q_N$ is the AMOC strength from (\ref{eq:q_n}).
Since a continuous step-like change  of the AMOC  is found in the CESM (see Figure~4 in 
\citet{vanWesten2023c}),  we use a similar  function for the reduction factor 
(Figure~\ref{fig:Figure_A2}b). We choose 
\begin{equation} \label{eq:q_n_min}
r_q     = \frac{1-r_q^{\mathrm{min}}}{2} \tanh(0.1 \times (0.5 - f_\mathrm{n})) +  \frac{1+r_q^{\mathrm{min}}}{2} . 
\end{equation}
such that  $r_q \sim 1$ when the sea-ice fraction $f_\mathrm{n} < 0.4$ (40\%) and, for $f_\mathrm{n} > 0.6$ (60\%), $r_q$ 
decreases  to a minimum value $r_q^{\mathrm{min}}$. When $r_q =  1$ is chosen, the model does not represent any 
sea ice effects. 

In  the CESM results \citep{vanWesten2023c},  there is a factor 10~difference in spatially-averaged 
mixed layer  depth over the Irminger basin  between the two AMOC states (which would suggest 
$r_q^{\mathrm{min}} \approx 0.1$) with local differences up to a factor  20~difference 
($r_q^{\mathrm{min}} \approx 0.05$). We will vary $r_q^{\mathrm{min}}$ below to investigate its sensitivity in the the E-CCM.
Note that the value of  $r_q$ depends on the sea-ice fraction 
and hence on the temperature $T_\mathrm{n}$, but sea ice is not dynamically resolved in the E-CCM 
and its effect is only parameterised. 

\subsection{Steady States and Transition Probabilities}

To determine the bifurcation diagrams of the CCM and E-CCM we implemented the model equations in the continuation 
software AUTO-07p \citep{Doedel2007,Doedel2021}. The AUTO software is able to directly compute steady states 
versus model parameters   without time integration by using a pseudo-arclength continuation combined with a 
Newton-Raphson method.  An additional advantage of AUTO is its ability to detect special points such as 
Hopf and saddle-node  bifurcations, which are both present in the CCM and E-CCM. 
The accuracy of AUTO is determined by three coefficients  representing the absolute and relative accuracy 
of the solution, and the accuracy for locating special points. All these coefficients  are set on a minimum value 
of 10$^{-6}$ and a higher accuracy is used where necessary to determine an accurate bifurcation diagram.

In the multi-stable regime, we study  noise-induced transitions in the stochastic version 
of the  E-CCM. Stochastic variations in the asymmetric freshwater flux forcing
are considered by adding a term $f_{\sigma} E_A S_0 \zeta(t)$ to the salinity equation
of box~n (and subtracting it from the salinity equation of box~s), where $\zeta(t)$ is a zero 
mean, delta time correlated
(white) noise process and  $f_{\sigma}$ controls the strength of the noise.  Following 
\citet{Castellana2019}, we use  the Trajectory Adaptive Multilevel 
Splitting (TAMS)  algorithm \citep{Lestang2018}  to determine  probabilities of transitions 
between  different steady AMOC states in the E-CCM within a given time 
interval $T$.  

\section{Results}

\subsection{Statistical Steady States in CESM}

\begin{figure}

\vspace{-2cm} 

\center
\includegraphics[width=1\columnwidth]{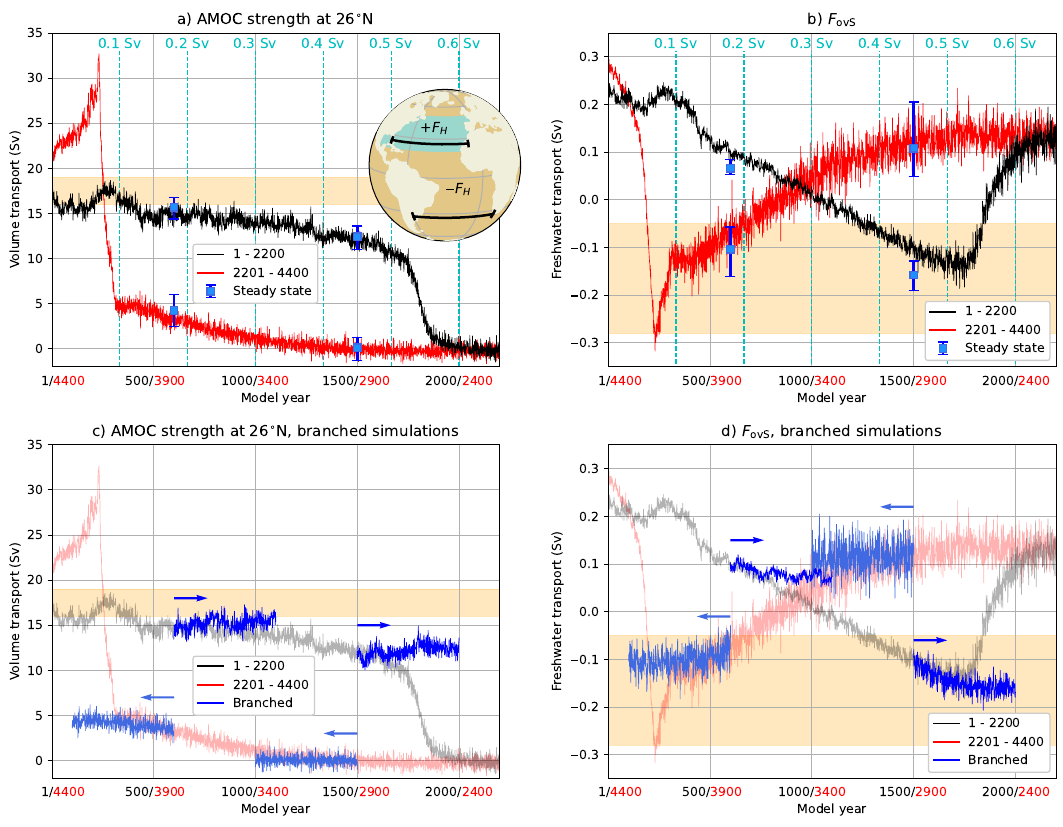}
\caption{(a): The AMOC strength at 1,000~m and 26$^{\circ}$N.
The cyan-coloured lines in panels~a and b indicate the magnitude of the freshwater forcing $F_H$
and the yellow shading in all panels indicates observed ranges for the presented quantity \citep{Garzoli2013, Mecking2017, Smeed2018, Worthington2021}.
The statistical equilibria (i.e., steady states in blue, panels~a and b) of model years~600, 1,500, 2,900 and 3,800 are also shown,
where the marker indicates the mean and the error bars show the minimum and maximum over the last 50~years of the 600-year long branched simulations.
Inset: The hosing experiment where fresh water is added to the ocean surface between 20$^{\circ}$N -- 50$^{\circ}$N in the Atlantic Ocean ($+F_H$) 
and is compensated over the remaining ocean surface ($-F_H$).
The black sections indicate the 26$^{\circ}$N and 34$^{\circ}$S latitudes over which the AMOC strength and $F_{\mathrm{ovS}}$ are determined, respectively.
(b): The freshwater transport by the AMOC at 34$^{\circ}$S, $F_{\mathrm{ovS}}$.
(c \& d): Similar to panels~a,b, but now the entire branched simulations are displayed.
The arrows indicate the beginning of the branches and the direction of time for each branch.}
\label{fig:Figure_2}
\end{figure}

The branched simulations from the hysteresis experiment identify multiple statistical 
steady AMOC states in the CESM (Figure~\ref{fig:Figure_2}) for both $F_H = 0.18$
~Sv and $F_H = 0.45$~Sv.   

The first state is obtained from the strong northward overturning state for $F_H = 0.18$~Sv 
(branched from model year~600).  The AMOC strength appears statistically stationary  after 500~years 
of the branched simulation.  When we consider the time means (450 -- 500~years after the branching 
point),  we find an AMOC strength and $F_{\mathrm{ovS}}$ of 15.7~Sv and 0.07~Sv, respectively.
The results from the quasi-equilibrium simulation (model years~575 -- 624) are close to the latter 
equilibrium values with 14.8~Sv for AMOC strength and 0.10~Sv for $F_{\mathrm{ovS}}$. 
For $F_H = 0.18$~Sv (branched from model year~3,800), there is a second statistical  steady state.
This state has a weak and shallow ($<$1,000~m)  northward overturning cell \citep{vanWesten2023c}
in the Atlantic Ocean with an equilibrium AMOC strength of 4.2~Sv and  $F_{\mathrm{ovS}}$ of $-$0.10~Sv.
The quasi-equilibrium values (model years~3,775 -- 3,824) are 3.2~Sv and $-$0.07~Sv for the 
AMOC strength and $F_{\mathrm{ovS}}$, respectively.  

For $F_H = 0.45$~Sv (branched from model year~1,500), the AMOC is still in the northward 
overturning regime, but its strength has declined under the stronger freshwater forcing. 
The equilibrium values are 12.4~Sv for AMOC strength and $-$0.16~Sv for $F_{\mathrm{ovS}}$,
while the quasi-equilibrium approach (model years~1,475 -- 1,524) values are 11.9~Sv and $-$0.10~Sv, respectively.
As discussed in \citet{vanWesten2024}, the sign and variance in $F_{\mathrm{ovS}}$ are good 
indicators of the approach of the AMOC tipping point. The $F_{\mathrm{ovS}}$ variance for $F_H = 
0.45$~Sv ($2.1 \times 10^{-4}$~Sv$^2$) is indeed larger than at $F_H = 0.18$~Sv ($0.33 \times 
10^{-4}$~Sv$^2$), which is directly visible from the branched time series (Figure~\ref{fig:Figure_2}d).
A linear trend was removed before determining the variances. The larger variance indicates that the 
AMOC loses resilience and is closer to its tipping point. The variances in AMOC strength are quite 
similar for both branches and are about 0.3~Sv$^2$. 
For $F_H = 0.45$~Sv another statistical steady state exists; this is the collapsed AMOC state 
(branched from model year~2,900).  This equilibrium state has an AMOC strength of 0.08~Sv and  
$F_{\mathrm{ovS}}$ of 0.11~Sv  and the values for the quasi-equilibrium approach (model 
years~2,875 -- 2,924) are close to the equilibrium values. 

The quasi-equilibrium simulation  remains reasonably close to the steady states but also indicates that 
the freshwater forcing changes are faster than oceanic adjustment time scales.
Ideally one would like to change the freshwater forcing even slower (than $3 \times 10^{-4}$~Sv~yr$^{-1}$) and 
determine more steady states, preferably closer to the tipping points,
but these simulations require even more computational resources and are at the moment not feasible. 
As argued in \citet{vanWesten2023c}, one would expect the AMOC recovery to occur at the 
value of $F_H$ where $F_{\mathrm{ovS}}$ changes sign on the AMOC on 
branch (black curve in Figure~\ref{fig:Figure_2}b), that is around year~3,400 in the
quasi-equilibrium approach. The results in Figure~\ref{fig:Figure_2} show instead that a statistical 
steady AMOC off state still exists for $F_H = 0.18$~Sv and hence the late recovery at year 4,100 
is not due to an overshoot but to the fact that, in a model with sea ice, AMOC off  states 
can apparently extend to positive values of $F_{\mathrm{ovS}}$. We will explore this
below  in  more detail using the conceptual model. 

\subsection{Steady States in the E-CCM}

\begin{figure}[h]

\includegraphics[width=1\columnwidth]{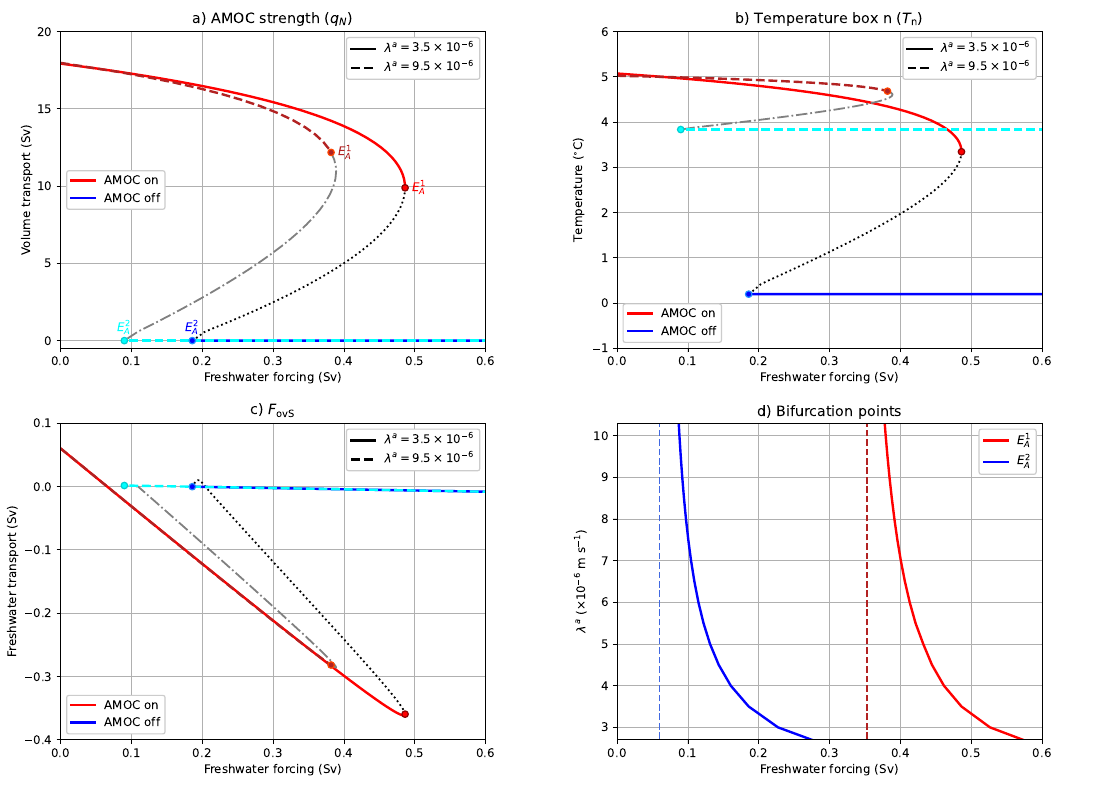}

\caption{(a -- c): The steady state solutions for the E-CCM (no sea ice) for the (a): AMOC strength, (b): 
temperature of box~n and (c): $F_{\mathrm{ovS}}$.  The red and blue curves indicate the 
steady AMOC on and AMOC off state, respectively, and for two different values for $\lambda^a$ 
(solid and dashed lines). The (dash)dotted curves are the unstable branches and the saddle-node bifurcation 
points are indicated by the circled markers. (d): The bifurcation points as a function of freshwater 
forcing for varying $\lambda^a$, the dashed lines indicate the bifurcation points for the CCM.}

\label{fig:Figure_3}
\end{figure}

To investigate the important role of sea ice in the AMOC hysteresis,  we compute  steady states in 
the deterministic E-CCM.  First we analyse the case without sea ice (by setting $r_q^{\mathrm{min}} 
= 1$) and vary the asymmetric freshwater flux forcing ($E_A$).  The bifurcation diagrams for this case 
are shown in Figure~\ref{fig:Figure_3}a for two values of $\lambda^a$. Both bifurcation points are saddle 
node bifurcations, similar to the CCM, and close to the saddle node bifurcations ($\Delta E_A = 0.002$~Sv) there are (subcritical) Hopf bifurcations which 
give unstable periodic orbits (not shown). The position of the  bifurcation points is dependent 
on the heat exchange rate $\lambda^a$ with the atmosphere and lower values of $\lambda^a$ (less restoring) result in 
a larger value of $E_A$ at the saddle-node bifurcation points. In the limit of $\lambda^a \rightarrow \infty$, 
we find the same bifurcation points as in the CCM  (dashed lines in Figure~\ref{fig:Figure_3}d). 
The hysteresis width, that is the distance between $E_A^1$ and $E_A^2$,  is about 0.3~Sv and does not 
depend much on  $\lambda^a$.

To understand this behaviour of the shift of $E^1_A$ with $\lambda^a$, we analyse the AMOC strength response under varying $E_A$.
The AMOC strength is represented by $q_N$ and is
proportional to the density difference between box~n and box~ts ($\Delta \rho$),
and, given the linear equation of state, it boils down to $\Delta \rho \sim \Delta T - \Delta S$.
Under the increasing freshwater forcing the salinity difference becomes larger and this decreases
the density difference resulting in AMOC weakening (Figure~\ref{fig:Figure_3}a).
The temperature of box~n ($T_\mathrm{n}$, Figure~\ref{fig:Figure_3}b) 
decreases as this box receives less heat from box~t under AMOC weakening, which enhances the 
temperature difference between box~n and box~ts (the temperature of box~ts remains fairly constant).
This larger temperature difference between box~n and box~ts increases the density difference, so there 
are two counteracting effects that control the density difference. The combined effect of the temperature 
and salinity responses under the freshwater forcing shifts the saddle-node bifurcation point ($E_A^1$, 
Figure~\ref{fig:Figure_3}d)  to higher values of freshwater forcing when comparing this to the 
standard CCM  (dashed lines in Figure~\ref{fig:Figure_3}d). The stabilising temperature effect is larger 
at smaller $\lambda^a$ (less restoring) and hence a larger value of $E^1_A$ is needed to destabilise 
the AMOC (Figure~\ref{fig:Figure_3}) at the saddle-node bifurcation. 

When the AMOC collapses, both temperature and salinity (strongly) decline in box~n, these changes are 
much smaller for box~ts. The density gradient is influenced by both salinity and temperature, the temperature 
response partly compensates the salinity response and   hence the  bifurcation point ($E_A^2$, 
Figure~\ref{fig:Figure_3}d) also shifts to higher values of $E_A$ compared to that of the CCM. 
Note that the temperature of box~n does not vary much  under the varying freshwater forcing 
when AMOC is in its off state (Figure~\ref{fig:Figure_3}b). This box is in equilibrium as it only 
receives heat from its overhead atmosphere and from the gyre transport with box~t. 
The Hopf bifurcations near $E_A^1$ and $E_A^2$ have associated periods of about 60 to 200~years and 3,000 to 6,000~years, respectively.
For relatively large values of $\lambda^a$,  the value of $E_A$ where  $F_{\mathrm{ovS}}$ changes sign
on the AMOC on branch is close to $E_A^2$, similar as in the CCM.
But for small values of $\lambda^a$ (e.g., $\lambda^a = 3.5 \times 10^{-6}$~m~s$^{-1}$), this is clearly not the case.
In other words, negative $F_{\mathrm{ovS}}$ does not exclusively indicate the multi-stable regime.

\begin{figure}[h]

\includegraphics[width=1\columnwidth]{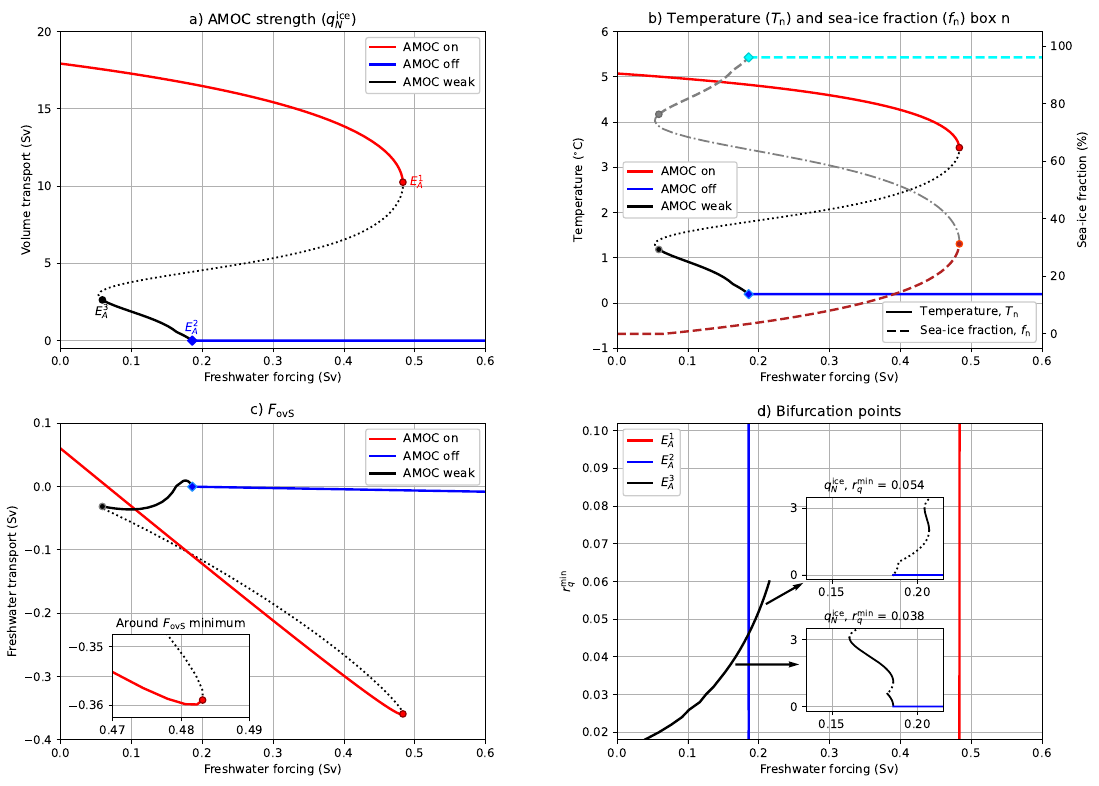}

\caption{(a -- c): The steady state solutions for the E-CCM for the (a): AMOC strength,
(b): temperature of box~n and sea-ice fractions and (c): $F_{\mathrm{ovS}}$. 
The red, blue and black curves indicate the AMOC on, AMOC off and AMOC weak state, respectively, 
for $\lambda^a = 3.5 \times 10^{-6}$~m~s$^{-1}$ and $r_q^{\mathrm{min}} = 0.02$.
The (dash)dotted curves are the unstable branches and the saddle-node bifurcation points are indicated by the circled markers,
the diamond marker is not a saddle-node bifurcation point.
The inset in panel~c is a zoom in around the $F_{\mathrm{ovS}}$ minimum of the AMOC on branch.
(d): The bifurcation points as a function of $E_A$  for varying $r_q^{\mathrm{min}}$ and for 
$\lambda^a = 3.5 \times 10^{-6}$~m~s$^{-1}$.
The two insets shows the steady state for AMOC strength ($q_N^{\mathrm{ice}}$, similar to panel~a) for
$r_q^{\mathrm{min}} = 0.038$ and $r_q^{\mathrm{min}} = 0.054$, centred around the weak state.}

\label{fig:Figure_4}
\end{figure}

Next, the case with sea-ice representation in the E-CCM is considered for 
$\lambda^a = 3.5 \times 10^{-6}$~m~s$^{-1}$.  Figure~\ref{fig:Figure_4} shows the bifurcation diagram for  
$r_q^{\mathrm{min}} = 0.02$, where the saddle node bifurcation $E^2_A$ in 
Figure~\ref{fig:Figure_3}  has moved to $E^3_A$.  The saddle-node $E_A^1$ is hardly 
affected as the AMOC reduction factor $r_q$ is about~1 (relatively low sea-ice fractions 
when $f_\mathrm{n} < 35\%$) when in the AMOC on state.  At the value  $E_A^2$ there is no longer 
a saddle-node bifurcation,  but it remains an important point as $q_N$ switches sign here; 
in the case without sea ice this would then result in AMOC recovery.  However, in the presence 
of sea ice, the AMOC reduction factor is very strong (here $r_q  \approx 0.02$) and a new branch of 
equilibrium states appears in which the AMOC remains weak (less than 3~Sv). For 
decreasing values of the freshwater forcing, the AMOC becomes stronger and the sea-ice 
fraction starts declining until the AMOC reaches about 3~Sv at the new bifurcation point $E_A^3$. 
There are again (subcritical) Hopf bifurcations close to the saddle nodes $E_A^1$ and $E_A^3$ with 
associated periods of 50 to 250~years and 200 to 1,250~years, respectively.

When decreasing the freshwater forcing from $E_A^2$, $F_{\mathrm{ovS}}$ becomes slightly positive and then changes sign.
In this freshwater forcing range,  the AMOC strength is very weak ($q_N^{\mathrm{ice}} < 0.6$~Sv) and $q_S$ remains negative.
Lowering the freshwater forcing further results in a stronger AMOC and $q_S$ becomes positive, which induces a negative $F_{\mathrm{ovS}}$. 
In the CCM, negative values of $F_{\mathrm{ovS}}$ in the AMOC on state mark the multi-equilibrium regime \citep{Castellana2019}
and here for $r_q^{\mathrm{min}}= 0.02$ this is (by coincidence) also the case (Figure~\ref{fig:Figure_4}c).
However, for $r_q^{\mathrm{min}}< 0.02$ the bifurcation point $E_A^3$ shifts to lower values of the freshwater forcing
and then positive $F_{\mathrm{ovS}}$ are in multi-equilibrium regime, similar as in the CESM.
The quantity $F_{\mathrm{ovS}}$ is still an important indicator in the CESM as its sign change
coincides with the initial AMOC recovery but the AMOC strength is strongly suppressed by sea-ice processes \citep{vanWesten2023c}.
When using a larger $\lambda^a$ and a different sea-ice response function in the E-CCM,
we can tune the model such that $E_A^2$ is close the $F_{\mathrm{ovS}}$ sign change, similar as in the CESM.

It is interesting that the weak state splits into two sub-branches for $0.030 \leq r_q^{\mathrm{min}} \leq 0.052$ (inset in Figure~\ref{fig:Figure_4}d for $r_q^{\mathrm{min}} = 0.038$);
the AMOC on and AMOC off states remain very similar.
The two sub-branches differ in $F_{\mathrm{ovS}}$ sign and there are subcritical Hopf bifurcations near three saddle-node bifurcations located on $E_A^1$, $E_A^2$, and $E_A^3$.
The associated periods for the Hopf bifurcations near $E_A^1$ and $E_A^3$ are 50 -- 500~years and 150 -- 500~years, respectively.
The Hopf bifurcation near $E_A^2$ has a much longer associated period of 44,000 -- 52,000~years.
For $0.052 < r_q^{\mathrm{min}} \leq 0.060$ only the weak state with negative $F_{\mathrm{ovS}}$ remains (inset in Figure~\ref{fig:Figure_4}d for $r_q^{\mathrm{min}} = 0.054$) 
and $E_A^2$ is now a saddle node (similar to Figure~\ref{fig:Figure_3}).
For increasing $r_q^{\mathrm{min}}$, the saddle node $E_A^3$ shifts to larger values of $E_A$  and has a maximum AMOC strength of 3~Sv.
The AMOC weak state only appears when the AMOC reduction factor is sufficiently low ($r_q^{\mathrm{min}} \leq 0.060$) and 
for very low values of this factor ($r_q^{\mathrm{min}} < 0.015$) a negative freshwater forcing is required to return to the AMOC on state (Figure~\ref{fig:Figure_4}d).
Note that for $r_q^{\mathrm{min}} > 0.060$ the bifurcation point $E_A^3$ disappears and then the results are similar to Figure~\ref{fig:Figure_3}.
The hysteresis width also varies for varying $r_q^{\mathrm{min}}$. 
For values of $r_q^{\mathrm{min}}  \geq 0.046$, it is a constant width of 0.3~Sv
while the hysteresis width increases for lower values for $r_q^{\mathrm{min}}$.
For example, for $r_q^{\mathrm{min}} = 0.02$ it is about 0.4~Sv which  is comparable to that of the CESM.

\subsection{Quasi-equilibrium Experiment in the E-CCM}

\begin{figure}
\center
\includegraphics[width=1\columnwidth]{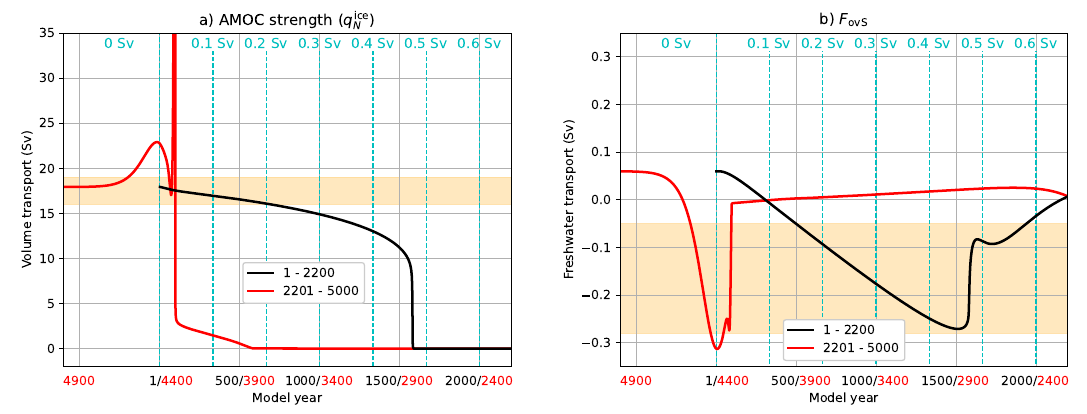}
\caption{(a): The AMOC strength ($q_N^{\mathrm{ice}}$) and (b): the freshwater transport by the AMOC at 30$^{\circ}$S in the E-CCM
for $\lambda^a = 3.5 \times 10^{-6}$~m~s$^{-1}$ and $r_q^{\mathrm{min}} = 0.02$.
The cyan-coloured lines in panels~a and b indicate the magnitude of $E_A$.
The yellow shading in all panels indicates observed ranges for AMOC strength and $F_{\mathrm{ovS}}$.
Note that after model year~4,400 the freshwater forcing is held constant at $E_A = 0$~Sv.
}
\label{fig:Figure_5}
\end{figure}

When varying the freshwater flux forcing over time with a rate of $3 \times 10^{-4}$~Sv~yr$^{-1}$ (similar to the CESM),
we find qualitatively the same hysteresis behaviour in the E-CCM (Figure~\ref{fig:Figure_5}) as in the CESM (Figure~\ref{fig:Figure_2}).
Note that we continued the simulation beyond model year~4,400 under no freshwater forcing to run the model into equilibrium. 
We find an AMOC tipping event around model year~1,600 and $F_{\mathrm{ovS}}$ goes through a minimum and further equilibrates to the reversed circulation. 
The steady states show the $F_{\mathrm{ovS}}$ minimum prior to the AMOC tipping point (inset in Figure~\ref{fig:Figure_4}c)
with a difference of only $\Delta E_A =  8 \times 10^{-4}$~Sv.
Other idealised models find (qualitatively) a similar timing between $F_{\mathrm{ovS}}$ minimum and AMOC tipping \citep{Dijkstra2007, Huisman2010}
and in the CESM \citep{vanWesten2024}.

The AMOC remains in its off state until around model year~3,900 and $q_N^{\mathrm{ice}}$ becomes positive and $F_{\mathrm{ovS}}$ switches sign. 
Eventually the AMOC transitions to the on state and strongly overshoots (to 68~Sv).
This overshoot is a model artefact as the AMOC strength is then determined by the (instantly felt) density difference between box~n and box~ts 
and by the pycnocline depth which is about twice as deep.
The AMOC in the CESM is not only controlled by this density difference and the pycnocline depth,
which explains why the CESM does not have such a large overshoot.
There is a second overshoot in AMOC strength of about 22.9~Sv (around model year~4,400),
which is more realistic and dynamically driven.
Advection of temperature and salinity controls the adjustment time scale from the weak AMOC to the AMOC on state and this takes about a few centuries, 
which is much faster than in the CESM (or real ocean), where vertical mixing also plays an important role.
Nevertheless, prior to the full adjustment to the AMOC on state, the different temperature and salinity gradients (w.r.t. the initial state) allow for a temporarily stronger AMOC strength.
At the same time, $F_{\mathrm{ovS}}$ reaches its minimum and then switches sign when the salinity in the boxes further adjust to the AMOC on state.

The transient responses in the E-CCM are comparable to that of the CESM.
The deviations in the quasi-equilibrium approach w.r.t. the steady states of the AMOC on branch become larger when approaching the tipping point.
Note that the timing of AMOC collapse and recovery is not identical to that of the CESM,
it is possible to vary the $\lambda^a$ and the AMOC reduction and sea-ice response functions to tune the E-CCM even better to the CESM results.
The important result here is that the E-CCM reasonably captures the AMOC responses from the CESM
with only a few adjustment to the original model.

\subsection{Transition Probabilities in the E-CCM}

We define an on-to-off transition as a transition from the steady AMOC on state to an AMOC state which strength 
is below a certain threshold (here taken as 0~Sv for the no sea-ice case and 1~Sv for the sea-ice case).  In 
addition, an off-to-on transition is a transition from  the steady AMOC off state to an AMOC which strength is larger or equal to the strength of the  steady AMOC on state. 
We repeat the TAMS procedure \citep{Lestang2018} 15~times and then determine the 
mean transition probability $p$ (and  its standard deviation)  of specific transitions within  $T = 100$~years, similar to \citet{Castellana2019}. Given this relatively short time interval (w.r.t. 
the dynamics of the E-CCM) we only find typical F-type (fast) transitions  \citep{Castellana2019}.
Full collapses (i.e., AMOC on $\rightarrow$ AMOC collapsed) or full recoveries (i.e., AMOC collapsed 
$\rightarrow$ AMOC on), i.e. S-type transitions    \citep{Castellana2019}, take much longer time intervals 
(a few centuries) and are not considered. 

\begin{figure}
\center
\includegraphics[width=1\columnwidth]{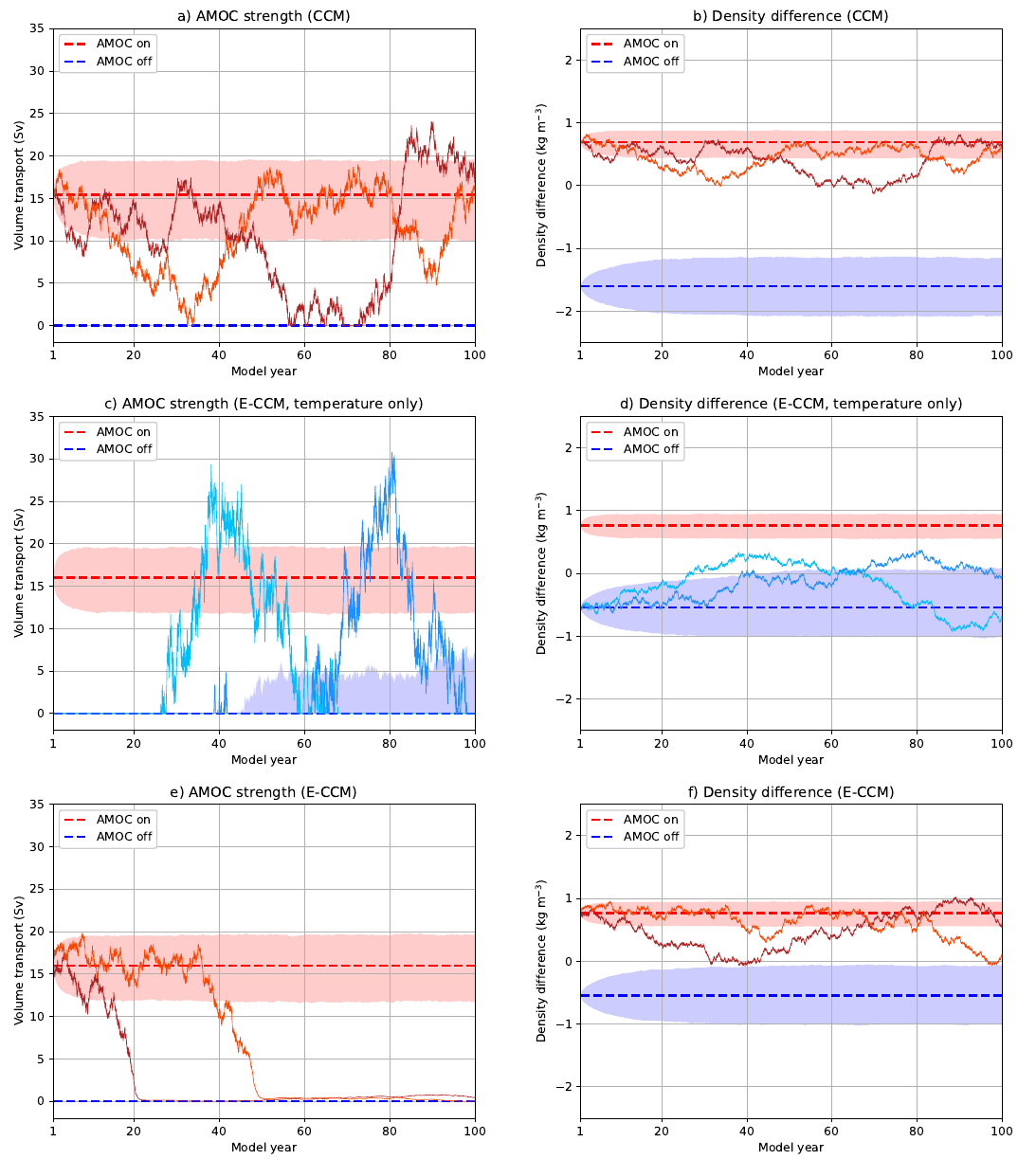}
\caption{Transitions in the (E-)CCM for $f_{\sigma} = 0.085$ and $E_A = 0.25$~Sv
for the AMOC strength (left column) and density difference between box~n and box~ts (right column).
The dashed lines indicate the statistical steady AMOC on (red) and the AMOC off (blue) statistical steady 
state, where the shading indicates the 95\%-confidence level derived from 5,000~different noise realisations 
(for each statistical steady state).  A few random realisations are displayed which (temporarily) undergo 
an on-to-off transition (red curves) or an off-to-on transition (blue curves).}
\label{fig:Figure_6}
\end{figure}

\begin{figure}
\center
\includegraphics[width=1\columnwidth]{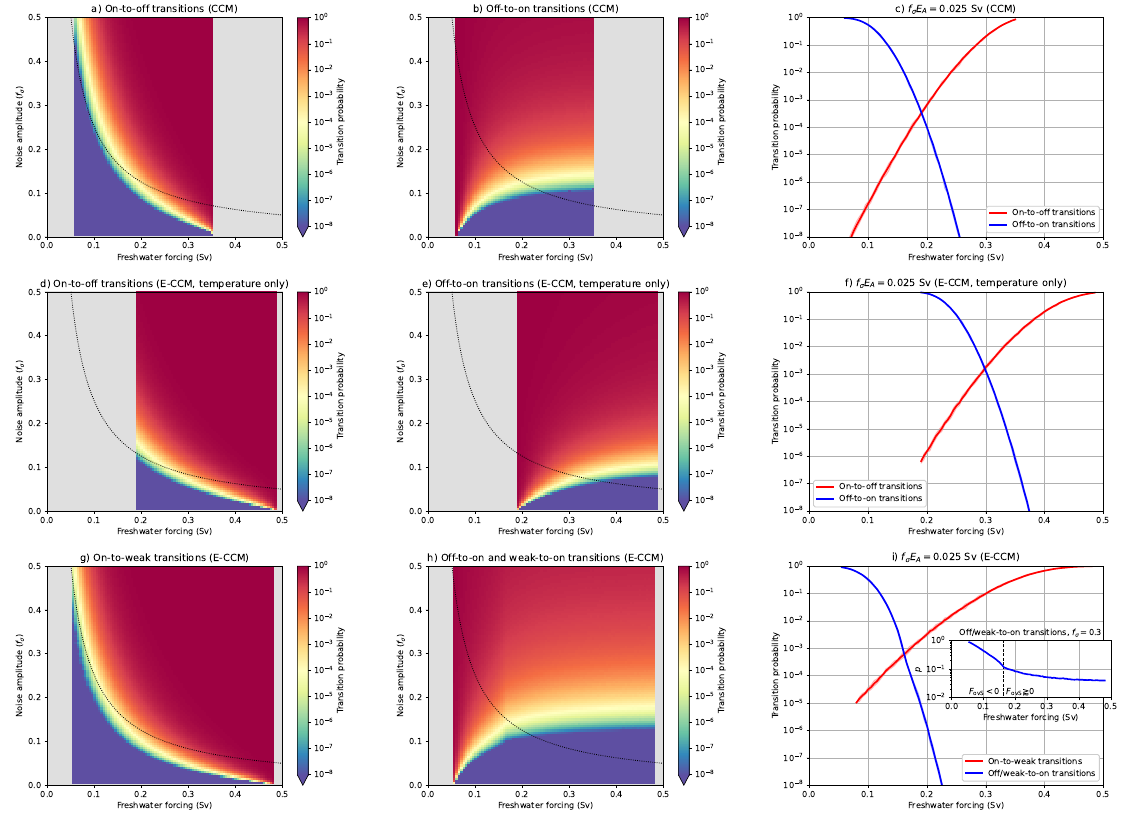}
\caption{Mean transition probabilities (within 100~years) in the (E-)CCM for varying $f_{\sigma}$ and $E_A$, 
with $\Delta f_{\sigma}$ = 0.005 and $\Delta E_A$ = 0.005~Sv.
The on-to-off (or on-to-weak) transitions are shown in the left column and 
the off(/weak)-to-on transitions in the middle column.
For the regions shaded in gray only one AMOC steady state exists.
The dotted curve indicates the freshwater noise forcing of 0.025~Sv (= $f_{\sigma} E_A$) and 
the mean transition probabilities along this freshwater noise forcing are shown in the right column,
the shading indicates the minimum and maximum transition probabilities derived from 
15~independent TAMS realisations. 
The inset in panel~i shows the mean transition probabilities for the E-CCM for $f_{\sigma} = 0.3$ under varying $E_A$,
and only for off/weak-to-on transitions.
}
\label{fig:Figure_7}
\end{figure}

The CCM forced by freshwater noise ($f_{\sigma} = 0.085$) for $E_A = 0.25$~Sv
undergoes an on-to-off transition and then quickly returns to its AMOC on state 
(Figures~\ref{fig:Figure_6}a). These AMOC transitions are reflected in the sign change 
of the density difference between box~n and box~ts (Figures~\ref{fig:Figure_6}b).
The AMOC state is controlled by the sign in density difference and its magnitude 
determines the transition probability. The magnitude of this density difference is larger 
in the steady AMOC off state ($-$1.6~kg~m$^{-3}$) than the steady AMOC on state (+0.7~kg~m$^{-3}$) 
and the larger the difference in absolute magnitude the lower the transition probabilities (Figures~\ref{fig:Figure_7}a,b).
In other words, realisations with large density differences require relatively large freshwater noise perturbation to make a transition.
The transition probability under this  freshwater noise forcing is $p < 10^{-9}$ for off-to-on transitions 
and is substantially lower than for on-to-off transitions ($p = 0.0025 \pm 0.0001$).
This much lower transition probability also explains why there are no off-to-on transitions for this particular noise realisation
as more than $10^9$ noise realisations are required to expect at least 1~transition, whereas we used only 5,000~realisations in Figures~\ref{fig:Figure_6}a,b. Lowering the freshwater forcing towards the second bifurcation point ($E_A^2$ $\approx$0.06~Sv) 
would favour off-to-on transitions  while on the other hand the probability of on-to-off transitions decreases 
(Figures~\ref{fig:Figure_7}c).

Under the same freshwater noise forcing for the E-CCM ($r_q = 1$ (no sea-ice effects), $\lambda^a = 3.5 \times 10^{-6}$~m~s$^{-1}$),
we mainly find off-to-on transitions (Figures~\ref{fig:Figure_6}c,d).
In this set-up the system is much closer to its second bifurcation point $E_A^2$; this is also reflected in the
smaller (absolute) density difference in the AMOC off state ($-$0.5~kg~m$^{-3}$) than in the AMOC on state (+0.8~kg~m$^{-3}$).
Under these freshwater noise forcing, the transition probabilities are $p = 7.9 \times 10^{-7} \pm 0.9 \times 10^{-7}$ (on-to-off 
transitions) and $p = 0.064 \pm 0.002$ (off-to-on transitions).
The overall transition probabilities in the E-CCM with no sea-ice effects (Figures~\ref{fig:Figure_7}d,e,f) are higher 
than in the CCM (Figures~\ref{fig:Figure_7}a,b,c)
when using the distance to the bifurcation points as a reference for both model configurations.
This is because the magnitude of the freshwater noise forcing is given by $f_{\sigma} E_A$ and, as 
the bifurcation points are shifted to higher values of $E_A$ (Figure~\ref{fig:Figure_3}d) when 
temperature is included, it results in a larger freshwater noise forcing allowing for higher transition 
probabilities.

The density differences remain unaltered ($<0.001\%$ difference w.r.t. $r_q = 1$) when including sea ice 
($r_q^{\mathrm{min}} = 0.02$) in the E-CCM. Yet, there are now on-to-off transitions (Figures~\ref{fig:Figure_6}e,f).
The density difference of the two (random) realisations remains positive after an AMOC collapse and the AMOC 
reduction factor (in the presence of sufficient sea ice) is strong enough to limit their AMOC strength to almost 0~Sv.
On relatively short time scales ($<$ 100~years) the system undergoes an F-type transition and remains in a 
weak AMOC state (i.e., AMOC strength slightly larger than 0~Sv). Note that this state is unstable for the 
imposed freshwater forcing conditions and when the system further equilibrates, it makes the full transition to 
the stable AMOC collapsed state.
These transitions are somewhat different compared to the other two model set-ups and we define on-to-off transitions when $q_N^{\mathrm{ice }} < 1$~Sv, 
and we label them as on-to-weak transitions for the E-CCM.
Sea ice suppresses the AMOC strength and this strongly limits off-to-on transitions ($p < 10^{-9}$) compared 
to on-to-weak transitions ($p = 0.0031 \pm 0.0003$) under $f_{\sigma} = 0.085$ and $E_A = 0.25$~Sv.
The much lower transition probabilities (Figures~\ref{fig:Figure_7}g,h,i) compared to the other model 
set-ups clearly show that the AMOC weak/off state is much more resilient than the AMOC on state.
On the other hand, the on-to-weak transitions are more likely when including sea ice (compare red curves 
in Figures~\ref{fig:Figure_7}f and \ref{fig:Figure_7}i for $E_A^2 < E_A < E_A^1$) and this indicates 
that sea-ice processes destabilise the AMOC on state.

For $r_q^{\mathrm{min}} = 0.02$ one stable AMOC weak state exists and this state has both positive and negative $F_{\mathrm{ovS}}$ values (Figure~\ref{fig:Figure_4}).
Salinity perturbations are amplified under negative $F_{\mathrm{ovS}}$ (i.e., the salt-advection feedback) 
and this part of the AMOC weak state is relatively unstable compared to the part with positive $F_{\mathrm{ovS}}$.
The quantity $\phi = \left | \frac{\partial \log p}{\partial E_A} \right |$ is indeed larger for $F_{\mathrm{ovS}} < 0$ (inset in Figure~\ref{fig:Figure_7}i) and for varying $f_{\sigma}$,
indicating that the salt-advection feedback destabilises this part of the weak state where $F_{\mathrm{ovS}} < 0$ and resulting in higher transition probabilities.

\begin{figure}
\center
\includegraphics[width=1\columnwidth]{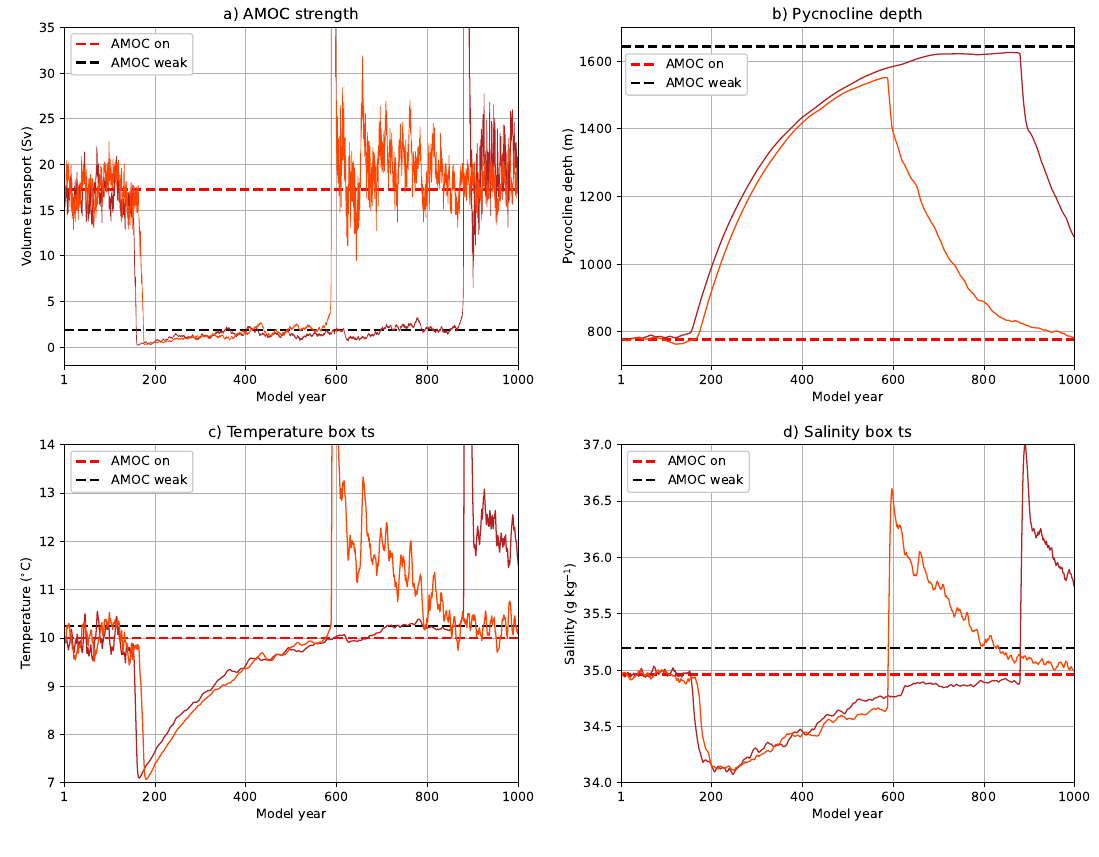}
\caption{Two random realisations undergoing transitions in the E-CCM ($\lambda^a = 3.5 \times 10^{-6}$~m~s$^{-1}$ and $r_q^{\mathrm{min}} = 0.02$) 
for $f_{\sigma} = 0.25$ and $E_A = 0.10$~Sv for various quantities.
The dashed lines indicate the steady AMOC on (red) and AMOC weak (black) states for each presented quantity.}
\label{fig:Figure_8}
\end{figure}

Once the E-CCM undergoes an on-to-weak transition, it is also possible to return back to the original AMOC on state.
Figure~\ref{fig:Figure_8} shows two realisations which undergo such behaviour for $f_{\sigma} = 0.25$ and $E_A = 0.10$~Sv.
Note that we simulate the trajectories for 1,000~years in Figure~\ref{fig:Figure_8}
because the model trajectory first needs to equilibrate to the AMOC weak state before the second transition is possible. 
The strong transient responses in the pycnocline and temperature and salinity for box~ts dominate the dynamics of the system 
as the AMOC strength is (partly) dependent on these oceanic properties.
The strong decline in the temperature and salinity (around model year~200) is related to the deepening of the pycnocline,
resulting in a volume expansion of box~ts and hence lowering of these quantities. 
The temperature and salinity in box~n (not shown) also decrease under an AMOC collapse and remain fairly constant from model year~200 until AMOC recovery.
Once the model trajectory  is sufficiently adjusted to the stable AMOC weak state (about 400~years after collapse), it  may undergo again random transitions to the AMOC on state.
The transition probabilities from Figure~\ref{fig:Figure_7} cannot (directly) be used to determine the transition probability of undergoing two consecutive transitions due to the setup of TAMS using fixed score functions \citep{Castellana2019}. 

\section{Summary and Discussion}

In this study we analysed AMOC hysteresis behaviour in the CESM  \citep{vanWesten2023c} and 
branched off four additional  simulations under constant freshwater forcing to determine  statistical steady 
states. We found three different statistical steady states in the CESM: the AMOC on state, the AMOC off state  
and the AMOC weak state. Sea-ice processes strongly influence the AMOC strength and modify the AMOC 
hysteresis \citep{vanWesten2023c}, giving rise to the AMOC weak state. This stable AMOC weak state 
is not found in ocean-only models in which sea-ice processes are not resolved \citep{Dijkstra2007, 
Huisman2010}. The existence of a broad multi-stable regime under varying freshwater forcing suggests 
that noise-induced AMOC tipping is possible in the CESM. However, we can not yet determine these 
transition probabilities for the CESM due to computational limitations. 

To determine the effects of the sea-ice processes on the AMOC transition probabilities, we 
developed a novel stochastic ocean-sea-ice box model (the E-CCM).   The original 5-box model
\citep{Cimatoribus2014, Castellana2019} was  extended by dynamically solving for temperatures and 
parameterising the effects of sea ice on the AMOC.  When the effects of sea ice are strong enough to 
suppress the AMOC strength, it modifies the AMOC hysteresis and gives rise to the AMOC weak state.
The deterministic box model can qualitatively reproduce the CESM hysteresis behaviour, which is 
again an example of the usefulness of conceptual models to understand AMOC dynamics 
\citep{Dijkstra2023b}.  The  existence of  this weak  state also illustrates the crucial importance of 
sea-ice processes on the AMOC strength \citep{vanWesten2023c}. This is in agreement with the results 
in \citet{Lin2023}, who show in CMIP6 models that sea-ice effects play a major role (through their 
effect on the Labrador Sea  stratification) in the decline of the AMOC under increased greenhouse 
gas concentrations. 

The sea-ice parametrisation in the E-CCM only affects the AMOC strength and does not consider brine rejection nor thermodynamics. 
Despite these shortcomings, Hopf bifurcations exist in the E-CCM and this is consistent with earlier works \citep{Yang1993, Brocker1990} 
where realistic sea-ice AMOC interactions have been studied to understand natural AMOC variability. 
Sea ice also plays an important role in Dansgaard-Oeschger events \citep{Peltier2014, Vettoretti2016, Boers2018}
and the recent AMOC hysteresis experiment conducted with the CESM confirms this \citep{vanWesten2023c}.
Biases in the North Atlantic sea-ice distribution influence the background AMOC strength and the
transient AMOC responses due to climate change  \citep{Lin2023} and can lead 
to rate-induced tipping \citep{Lohmann2021}. 
These studies and the results presented here suggest a prominent role for sea-ice processes in AMOC dynamics and variability.

The stochastic E-CCM enables to compute the transition probabilities between the different AMOC states 
under freshwater noise forcing using rare-event algorithms (i.e., TAMS, \citet{Lestang2018}). The presence 
of sea ice increases the probability of on-to-off and on-to-weak transitions and  sea ice strongly reduces 
the probability of weak-to-on and off-to-on transitions of the AMOC. This indicates that the AMOC off and weak 
states, which are both influenced by sea ice, are much more stable than the AMOC on state.  Although the
quantitative results will be sensitive to the parameterisation of sea ice and its effects on the AMOC in the 
E-CCM, they provide meaningful insights into the effects of sea-ice processes in the AMOC hysteresis 
behaviour and AMOC transition probabilities. 

An interesting follow-up study is to analyse the AMOC transition behaviour, and in  particular the role of sea  
ice,  under climate change. The suppressing effects of sea ice on the AMOC become smaller as sea ice is 
less likely to form under  higher temperatures. As the AMOC weak state is strongly controlled by sea ice, this 
state would (partly) disappear. This stabilises the AMOC as transition probabilities from the AMOC on to 
AMOC weak state  cease to exist over an interval of freshwater forcing.  On the other hand, climate change 
also induces different temperature gradients which may  destabilise the  AMOC. A combination of E-CCM 
analyses together with targeted CESM simulations can be very useful for such a follow-up study.

\clearpage
\acknowledgments
We thank Michael Kliphuis (IMAU, UU) for  performing  the additional CESM simulations.
The model simulations and the analysis of all the model output was conducted on 
the Dutch National  Supercomputer (Snellius) within NWO-SURF project~17239. 
R.M.v.W., A.A.B. and H.A.D. are funded by the European Research Council through the 
ERC-AdG project TAOC (PI: Dijkstra, project~101055096).  
V.J.D. is funded by the European Union’s Horizon 2020 research and innovation programme 
CriticalEarth under the Marie Sklodowska-Curie grant agreement (project~956170).
The authors declare no conflict of interest.

%
%
\datastatement
The Python codes for the E-CCM can be accessed through https://doi.org/10.5281/zenodo.10554659,
together with the CESM model output.








%



\appendix

\appendixtitle{Supporting Information on the Idealised Model Simulations}

The CCM consists of five equations for salinity (four dynamical and one to conserve salinity) and one equation for the pycnocline.
The equations are repeated here but we refer to \citet{Castellana2019} where all the different terms are explained ($\theta$ is the Heaviside function) and input parameters are also provided:

\begin{subequations} \label{eq:Box_model_salt}
\begin{align}
       \frac{\mathrm{d}(V_\mathrm{t} S_\mathrm{t})}{\mathrm{d}t} &= q_S \left(\theta(q_S)S_\mathrm{ts} + \theta(-q_S) S_\mathrm{t} \right) + q_U S_\mathrm{d} - \theta(q_N) q_N S_\mathrm{t} + r_S (S_\mathrm{ts} - S_\mathrm{t}) + r_N (S_\mathrm{n} - S_\mathrm{t}) + 2 E_S S_0 	 \\
       \frac{\mathrm{d}(V_\mathrm{ts} S_\mathrm{ts})}{\mathrm{d}t} &= q_{Ek} S_\mathrm{s} - q_e S_\mathrm{ts} - q_S \left(\theta(q_S)S_\mathrm{ts} + \theta(-q_S) S_\mathrm{t} \right) + r_S (S_\mathrm{t} - S_\mathrm{ts})  \\	
       \frac{\mathrm{d}(V_\mathrm{n} S_\mathrm{n})}{\mathrm{d}t} &= \theta(q_N) q_N \left(S_\mathrm{t} - S_\mathrm{n} \right) + r_N (S_\mathrm{t} - S_\mathrm{n}) - (E_S + E_A) S_0	\\
       \frac{\mathrm{d}(V_\mathrm{s} S_\mathrm{s})}{\mathrm{d}t} &= q_S \left( \theta(q_S) S_\mathrm{d} + \theta(-q_S) S_\mathrm{s} \right) + q_e S_\mathrm{ts} - q_{Ek} S_\mathrm{s} - (E_S - E_A) S_0 \\
        V_\mathrm{d} S_\mathrm{d} &=  S_0 V_0 - V_\mathrm{t} S_\mathrm{t} - V_\mathrm{ts} S_\mathrm{ts} - V_\mathrm{n} S_\mathrm{n} -  V_\mathrm{s} S_\mathrm{s} \\
       \left(A + \frac{L_{xA}L_{y}}{2} \right) \frac{\mathrm{d}D}{\mathrm{d}t} &= q_U + q_{Ek} - q_e - \theta(q_N) q_N
\end{align}
\end{subequations}

The ocean heat content evolution equations (under the assumption of constant heat capacity and reference density) are fairly similar to the salinity ones and are given by:

\begin{subequations} \label{eq:Box_model_temp}
\begin{align}
       \frac{\mathrm{d}(V_\mathrm{t} T_\mathrm{t})}{\mathrm{d}t} &= q_S \left(\theta(q_S)T_\mathrm{ts} + \theta(-q_S) T_\mathrm{t} \right) + q_U T_\mathrm{d} - \theta(q_N) q_N T_\mathrm{t} + r_S (T_\mathrm{ts} - T_\mathrm{t}) + r_N (T_\mathrm{n} - T_\mathrm{t}) - \lambda^a A_\mathrm{t} (T_\mathrm{t} - T_\mathrm{t}^a)	 \\
       \frac{\mathrm{d}(V_\mathrm{ts} T_\mathrm{ts})}{\mathrm{d}t} &= q_{Ek} T_\mathrm{s} - q_e T_\mathrm{ts} - q_S \left(\theta(q_S)T_\mathrm{ts} + \theta(-q_S) T_\mathrm{t} \right) + r_S (T_\mathrm{t} - T_\mathrm{ts})  - \lambda^a A_\mathrm{ts} (T_\mathrm{ts} - T_\mathrm{ts}^a) \\	
       \frac{\mathrm{d}(V_\mathrm{n} T_\mathrm{n})}{\mathrm{d}t} &= \theta(q_N) q_N \left(T_\mathrm{t} - T_\mathrm{n} \right) + r_N (T_\mathrm{t} - T_\mathrm{n})  - \lambda^a A_\mathrm{n} (T_\mathrm{n} - T_\mathrm{n}^a)	\\
       \frac{\mathrm{d}(V_\mathrm{s} T_\mathrm{s})}{\mathrm{d}t} &= q_S \left( \theta(q_S) T_\mathrm{d} + \theta(-q_S) T_\mathrm{s} \right) + q_e T_\mathrm{ts} - q_{Ek} T_\mathrm{s} - \lambda^a A_\mathrm{s} (T_\mathrm{s} - T_\mathrm{s}^a) \\	
       \frac{\mathrm{d}(V_\mathrm{d} T_\mathrm{d})}{\mathrm{d}t} &= \theta(q_N) q_N T_\mathrm{n} - q_S \left( \theta(q_S) T_\mathrm{d} + \theta(-q_S) T_\mathrm{s} \right) - q_U T_\mathrm{d}
\end{align}
\end{subequations}

Using the standard input parameters from \citet{Castellana2019}, we set the ocean surface areas to
$3 \times 10^{13}$~m$^{2}$ ($A_\mathrm{s}$), $1 \times 10^{13}$~m$^{2}$ ($A_\mathrm{ts}$), $10 \times 10^{13}$~m$^{2}$ ($A_\mathrm{t}$) and $1 \times 10^{13}$~m$^{2}$ ($A_\mathrm{n}$).
For more details how to integrate the model we refer to the publicly-available Python software.

\begin{figure}
\center
\includegraphics[width=1\columnwidth]{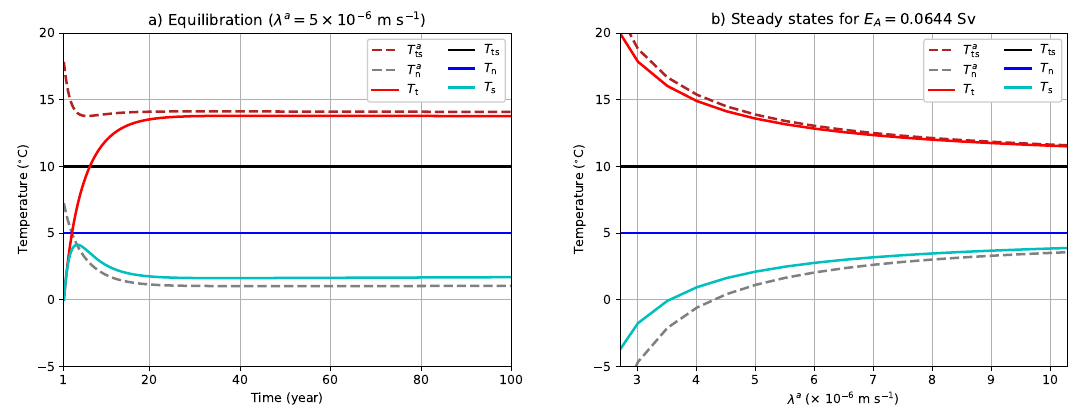}
\caption{(a): Evolution of the ocean (solid) and atmosphere (dashed) temperatures 
over the first 100~years for $\lambda^a = 5 \times 10^{-6}$~m~s$^{-1}$, where $T_\mathrm{n} = 5^{\circ}$C and $T_\mathrm{ts} = 10^{\circ}$C are fixed. 
(b): The steady states (after 2,000~years of integration) temperatures for varying $\lambda^a$.}
\label{fig:Figure_A1}
\end{figure}

To initialise the E-CCM,
we use the steady state solution from the CCM for an arbitrary freshwater forcing $E_A$ (here 0.0644~Sv) in the AMOC on state.
The temperatures for box~n and box~ts are initialised at 5$^{\circ}$C and 10$^{\circ}$C \citep{Castellana2019}, respectively, 
and the temperatures for the other boxes can be arbitrarily chosen (here 0$^{\circ}$C).
To remain at this steady state for the E-CCM, 
we solve for each time step the equations $\frac{\mathrm{d}(V_\mathrm{ts} T_\mathrm{ts})}{\mathrm{d}t} = 0$ and $\frac{\mathrm{d}(V_\mathrm{n} T_\mathrm{n})}{\mathrm{d}t} = 0$ for $T_\mathrm{ts}^a$ and $T_\mathrm{n}^a$, respectively.
This guarantees that the temperatures for box~ts and box~n remain the same over the entire simulation and hence we remain at the same steady state as the one from \citet{Castellana2019}.
To reduce the degrees of freedom, we assume the same atmospheric temperatures for the tropical boxes ($T_\mathrm{t}^a = T_\mathrm{ts}^a$) and polar boxes ($T_\mathrm{s}^a = T_\mathrm{n}^a$).
All the other temperatures are free to evolve and we quickly reach a steady state for a given $\lambda^a$ (Figure~\ref{fig:Figure_A1}a) and running the simulation further (2,000~years) does not change the steady states (Figure~\ref{fig:Figure_A1}b).
The steady states of the deep ocean temperature (not shown) are very close to $T_\mathrm{n}$ ($T_\mathrm{s}$) in the AMOC on state (off state).

Box~n receives relatively warm water from box~t (in the AMOC on state),
which has to be balanced by a colder overhead atmosphere so that box~n remains at a constant temperature of 5$^{\circ}$C.
For box~ts, it is the exact opposite, this box receives relatively cold water from box~s (in the AMOC on state) and a warmer overhead atmosphere is needed to balance this (negative) heat transport from box~s.
The temperature of box~s and box~t are also influenced by the changing atmospheric temperatures (they are coupled: $T_\mathrm{t}^a = T_\mathrm{ts}^a$ and $T_\mathrm{s}^a = T_\mathrm{n}^a$)
and remain close to their overhead atmospheric temperatures.
Decreasing $\lambda^a$ to values below $3 \times 10^{-6}$~m~s$^{-1}$ results in very cold (and unrealistic) atmospheric and oceanic temperatures (Figure~\ref{fig:Figure_A1}b).
In the limit of $\lambda^a \rightarrow \infty$, the atmospheric temperatures are 5$^{\circ}$C ($T_\mathrm{n}^a$) and 10$^{\circ}$C ($T_\mathrm{ts}^a$)
and temperature anomalies in box~n and box~ts decay immediately, resulting in constant temperatures of 5$^{\circ}$C and 10$^{\circ}$C, respectively.
This gives the same results as in the CCM.
After determining the steady states for a given $\lambda^a$ (Figure~\ref{fig:Figure_A1}b), we fix the atmospheric temperatures and use those as input parameters for the E-CCM.
All ocean temperatures are now free to evolve under a changing freshwater forcing $E_A$ and  the steady state is identical to the one from \citet{Castellana2019}
for the reference $E_A$ value (here 0.0644~Sv).


\begin{figure}
\center
\includegraphics[width=1\columnwidth]{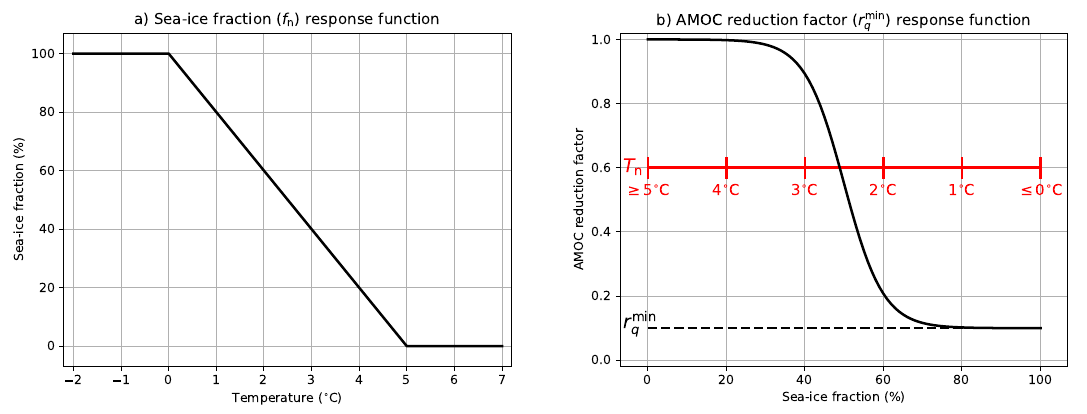}
\caption{(a): Sea-ice response function as a function of the temperature of box~n. 
(b): The AMOC reduction factor as a function of the sea-ice fraction of box~n for a given $r_q^{\mathrm{min}}$.}
\label{fig:Figure_A2}
\end{figure}



\bibliographystyle{ametsocV6}

\end{document}